\documentclass{article}
\usepackage{slashed,feynmf,epsfig,amsmath,amssymb,enumitem}
\usepackage[papersize={8.5in,11in}]{geometry}
\begin{document}
\title{Bimetric Gravity, Variable Speed of Light Cosmology and Planck2013}
\author{J. W. Moffat\\~\\
Perimeter Institute for Theoretical Physics, Waterloo, Ontario N2L 2Y5, Canada\\
and\\
Department of Physics and Astronomy, University of Waterloo, Waterloo,\\
Ontario N2L 3G1, Canada}
\maketitle

%\date{\today}


% ----------------------------------------------------------------

\begin{abstract}
A bimetric gravity model with a variable speed of light is shown to be in agreement with the results reported from the Planck satellite in 2013. The predicted scalar mode spectral index is $n_s\approx 0.96$ and its running is $\alpha_s\approx 8\times 10^{-4}$ when the fundamental length scale $\ell_0$ in the model is fixed to be $\ell_0\approx 10^5\ell_P$, where $\ell_P$ is the Planck length $\ell_P=1.62\times 10^{-33}\,{\rm cm}$, giving the observed CMB fluctuations: $\delta_H\approx 10^{-5}$. The enlarged lightcone ensures that horizon and flatness problems are solved. The model is free from many of the fine-tuning problems of the inflationary models and the fluctuations that form the seeds of structure formation do not lead to a chaotic inhomogeneous universe and the need for a multiverse. 
\end{abstract}

\maketitle

The Planck mission satelilite data~\cite{PlanckMission1,PlanckMission2,PlanckMission3} has disfavoured a wide class of inflationary models. The surviving inflation models are of the plateau-like models which, according to Ijjas, Steinhardt and Loeb~\cite{SteinhardtLoeb}, suffer from problems. They produce the unlikely possibility of allowing an initial inflationary phase and due to a chaotic post-inflationary evolution, the fluctuations responsible for primordial structure growth lead to a multiverse unpredictability (eternal inflation). After the Planck2013 data the inflationary paradigm faces serious difficulties, encouraging consideration of alternatives to inflation. Two possible alternatives are the periodic cyclic models with a bounce~\cite{SteinhardtTurok1,SteinhardtTurok2,SteinhardtLehners} and the variable speed of light (VSL) cosmology~\cite{Moffat:1993,Albrecht+Maguiejo:1999,Magueijo:2003,Magueijo+Moffat:2007}. In the following, we will investigate a version of VSL called bimetric gravity~\cite{Clayton+Moffat:1999,Clayton+Moffat:2000,Drummond:2001,Clayton+Moffat:2001,Clayton+Moffat:2002,Clayton+Moffat:2003a,Clayton+Moffat:2003b,Moffat:2003,Magueijo:2009,MagueijoPiazza:2010}.

A bimetric gravity theory was developed based on the geometric relationship:
\begin{equation}\label{eq:bimetric}
\hat {g}_{\mu\nu}=g_{\mu\nu}+B\partial_\mu\phi\partial_\nu\phi,
\end{equation}
with $B$ a constant with the dimensions of a squared length: $B=a\ell_0^2$ where $a$ is constant, and  $\phi$ is the biscalar field. The matter metric tensor ${\hat g}_{\mu\nu}$ provides the geometry on which matter fields propagate and interact, whereas the gravitational metric $g_{\mu\nu}$ represents the gravitational geometry through which gravitational waves and light waves propagate.
The presence of this type of prior-geometric structure in spacetime implies a differing propagation speed for matter, gravitational waves and light, and alters the coupling of matter to geometry. An alternative bimetric theory has also been developed based on the structure: $\hat {g}_{\mu\nu}=g_{\mu\nu}+B\psi_\mu\psi_\nu$ where $\psi_\mu$ is a vector field~\cite{Clayton+Moffat:1999}.

In contrast to the picture described by Hollands and Wald~\cite{Hollands+Wald:2002a,Hollands+Wald:2002b}, the radiation density in the volume of space determined by $\ell_0$ is highly diluted by means of a VSL
mechanism. This justifies the use of the biscalar field $\phi$ to generate the seeds of structure formation. The other important observation is that $\ell_0$ can be chosen to be related to the parameter $B$ in the bimetric structure through: $\ell_0\approx \sqrt{12B}$, and we obtain the correct order of magnitude of the CMB spectrum $\delta_H\approx 10^{-5}$ if $\ell_0\approx 10^5\ell_P=1.62\times 10^{-28}\,\mathrm{cm}$, where $\ell_P$ is the Planck length, $\ell_P=\sqrt{G\hbar/c^3}=1.62\times 10^{-33}\,\mathrm{cm}$. 

The model introduced in~\cite{Clayton+Moffat:1999,Clayton+Moffat:2000,Clayton+Moffat:2001,Clayton+Moffat:2002,Clayton+Moffat:2003a,Clayton+Moffat:2003b} consisted of a self-gravitating scalar field coupled to matter through the matter metric~\eqref{eq:bimetric}, with the action
\begin{equation}
S=S_{\rm grav}+S_{\phi}+\hat{S}_{\rm M},
\end{equation}
where
\begin{equation}
S_{\rm grav}=-\frac{1}{\kappa}\int d\mu (R[g_{\mu\nu}]+2\Lambda),
\end{equation}
$\kappa=16\pi G/c^4$, $\Lambda$ is the cosmological constant, and we employ a metric with signature $(+,-,-,-)$.  We will write, for
example, $d\mu=\sqrt{-g}\;d^4x$ and $\mu=\sqrt{-g}$ for the metric density related to the gravitational metric
$g_{\mu\nu}$, and similar definitions of $d\hat{\mu}$ and $\hat{\mu}$ in terms of the matter metric $\hat{g}_{\mu\nu}$, and $c$ denotes
the currently measured speed of light. Note that from~\eqref{eq:bimetric} the
determinants are related by
\begin{equation}\label{eq:density relation}
\mu=\sqrt{K}\hat{\mu},
\end{equation}
where
\begin{equation}\label{eq:K defn}
K=1-B\hat{g}^{\mu\nu}\partial_\mu\phi\partial_\nu\phi.
\end{equation}

The minimally-coupled scalar field action is
\begin{equation} S_{\rm \phi}=\frac{1}{\kappa}\int d\mu\,
\Bigl[\frac{1}{2}g^{\mu\nu}\partial_\mu\phi\partial_\nu\phi-V(\phi)\Bigr],
\end{equation}
where the scalar field $\phi$ has been chosen to be dimensionless.
The energy-momentum tensor for the scalar field that we will use is given by
\begin{equation}
T^{\mu\nu}_\phi=\frac{1}{\kappa}\Bigl[g^{\mu\alpha}g^{\nu\beta}\partial_\alpha\phi\partial_\beta\phi
-\tfrac{1}{2}g^{\mu\nu}g^{\alpha\beta}\partial_\alpha\phi\partial_\beta\phi
+g^{\mu\nu}V(\phi) \Bigr],
\end{equation}
and is the variation of the scalar field action with respect to the gravitational
metric: $\delta S_{\phi}/\delta g_{\mu\nu} =-\frac{1}{2}\mu T_\phi^{\mu\nu}$.

We construct the matter action $\hat{S}_{\mathrm{M}}$ using the
combination~\eqref{eq:bimetric} resulting in the identification of
$\hat{g}_{\mu\nu}$ as the physical metric that provides the arena on which
matter fields interact. That is, the matter action $\hat{S}_{\mathrm{M}}[\psi^I] =
\hat{S}_{\mathrm{M}}[\hat{g},\psi^I]$, where $\psi^I$ represents all the
matter fields in spacetime, is one of the standard forms, and therefore the
energy-momentum tensor derived from it by
\begin{equation}\label{eq:matterEM}
\frac{\delta S_{\mathrm{M}}}{\delta \hat{g}_{\mu\nu}}
 =-\frac{1}{2}\hat{\mu}\hat{T}^{\mu\nu},
\end{equation}
satisfies the conservation laws
\begin{equation}\label{eq:matterconservation}
\hat{\nabla}_\nu\Bigl[\hat{\mu}\hat{T}^{\mu\nu}\Bigr]=0,
\end{equation}
as a consequence of the matter field equations only~\cite{Clayton+Moffat:1999}.
It is the matter covariant derivative $\hat{\nabla}_\mu$ that appears here, which is
the metric compatible covariant derivative determined by the matter metric:
$\hat{\nabla}_\alpha\hat{g}_{\mu\nu}=0$. In this work we assume a perfect
fluid matter model:
\begin{equation}
 \hat{T}^{\alpha\beta}=
 \Bigl(\rho+\frac{p}{c^2}\Bigr)\hat{u}^\alpha\hat{u}^\beta
 -p\hat{g}^{\alpha\beta},
\end{equation}
with $\hat{g}_{\mu\nu}\hat{u}^\mu\hat{u}^\nu=c^2$.

The gravitational field equations for this model can be written as
\begin{equation}\label{eq:Einsteins eqns}
 G^{\mu\nu}=\Lambda g^{\mu\nu}
 +\frac{\kappa}{2}T^{\mu\nu}_\phi
 +\frac{\kappa}{2}\frac{\hat{\mu}}{\mu}\hat{T}^{\mu\nu}.
\end{equation}
The scalar field equation of motion is given by
\begin{equation}\label{eq:scalar FEQ}
\bar{g}^{\mu\nu}\hat{\nabla}_\mu\hat{\nabla}_\nu\phi+KV^\prime
[\phi]=0.
\end{equation}
In the latter, we have defined the biscalar field metric
\begin{equation}\label{eq:scalar metric}
 \bar{g}^{\mu\nu} = \hat{g}^{\mu\nu}
 +\frac{B}{K}\hat{\nabla}^\mu\phi\hat{\nabla}^\nu\phi
 -\kappa B\sqrt{K}\hat{T}^{\mu\nu},
\end{equation}
which is a third metric that controls the causal propagation of the biscalar field. We have previously shown that the field equations in this theory are consistent with the Bianchi identities~\cite{Clayton+Moffat:1999,Clayton+Moffat:2001}. Note that there is no source for the biscalar field in~\eqref{eq:scalar FEQ}. This implies that as the universe expands the biscalar field becomes increasingly diluted to the point where it has unobservable consequences at present, despite the rather large energy scale implied by the relation $\sqrt{12B}=\ell_0\sim 10^5\,\ell_P$ that we derive below.

We will work in a frame where the gravitational metric $g_{\mu\nu}$ is comoving:
\begin{equation}\label{eq:gravmetric}
{g}_{\mu\nu}=\mathrm{diag}(c^2,-a^2(t)\gamma_{ij}),
\end{equation}
with coordinates $(t, x^i)$ and $3$-metric $\gamma_{ij}$ on the spatial slices of constant time. As a result of this coordinate choice and the definitions~\eqref{eq:bimetric} and~\eqref{eq:scalar metric}, we find
\begin{equation}\label{eq:FRW ghat}
 \hat{g}_{\mu\nu}=\mathrm{diag}(K^{-1}c^2,-a^2(t)\gamma_{ij}),
\end{equation}
where from~\eqref{eq:K defn}:
\begin{equation}
 K = \left( 1 + \frac{B}{c^2}\dot{\phi}^2\right)^{-1}.
\end{equation}
We see from~\eqref{eq:gravmetric} that the constant $c$ will represent the speed of propagation of gravitational waves in this model, whereas from~\eqref{eq:FRW ghat} the speed of light (and the limiting speed of other matter fields) is time-dependent and given by
\begin{equation}
c_\gamma(t)=c/\sqrt{K}=c\left(1+\frac{B}{c^2}\dot{\phi}^2\right)^{1/2}.
\end{equation}
It follows that $c_\gamma(t)\rightarrow\infty$ and $K(t)\rightarrow 0$ when $B\dot\phi^2/c^2\rightarrow\infty$, and we expect that at present $K\sim 1$ and so $c_\gamma\sim c$.

We adopt an early universe radiation equation of state: $p=\frac{1}{3}c^2\rho$, and we will write:
\begin{equation}
\rho = \rho_{\mathit{rad},0}\left(\frac{a_0}{a}\right)^4,
\end{equation}
where the radiation energy density at present is: $\rho_{\mathit{rad},0}\approx 4.8\times 10^{-34}\,\mathrm{g}/\mathrm{cm}^3$.

The Friedmann equation is given by
\begin{equation}\label{eq:Fried eqn}
 H^2 +\frac{kc^2}{a^2} =
 \frac{1}{3}c^2\Lambda
+\frac{1}{6}\left(\frac{1}{2}\dot{\phi}^2+c^2V[\phi]\right)
 +\frac{1}{6}\kappa c^4 \sqrt{K}\rho,
\end{equation}
and the biscalar field equation is:
\begin{equation}\label{eq:biscalar eqn}
 (1-\kappa c^2 B K^{3/2}\rho)\ddot{\phi}
 +3H( 1+ \kappa B \sqrt{K}p)\dot{\phi}
 +c^2 V^\prime[\phi]
 =0,
\end{equation}
with the biscalar field metric given by:
\begin{equation}
 \bar{g}_{\mu\nu}
=\mathrm{diag}(c^2(1-\kappa c^2 B K^{3/2}\rho),
-a^2(t)(1+\kappa B \sqrt{K} p)\gamma_{ij}).
\end{equation}
The following quantities can be derived from the constant $B$:
\begin{equation}\label{eq:definitions}
H_B^2=\frac{c^2}{12B},\quad\rho_B=\frac{1}{2\kappa c^2 B},
\end{equation}
the latter comes from $H_B^2=\frac{1}{6}\kappa c^4\rho_B$.

We will now obtain the seeds for structure formation from a purely VSL mechanism by working in the comoving gravitational metric (\ref{eq:gravmetric}) (VSL metric frame). In the very early universe $K$ is very small: $B\dot{\phi}^2\gg c^2$, in which case the matter contribution to the Friedmann equation is small compared to the contribution from the biscalar field kinetic term. We are interested in a scenario that does \textit{not} depend on choosing a particular form for the biscalar field potential, and so we will set
$V[\phi]=0$.

We assume in~\eqref{eq:biscalar eqn} that the bracketed terms are both approximately one and so: $\bar{g}_{\mu\nu}\approx g_{\mu\nu}$, which gives us the solution:
\begin{equation}\label{eq:dot phi solution}
\dot{\phi}=\sqrt{12}H_B\left(\frac{a_{\mathit{pt}}}{a}\right)^3.
\end{equation}
We have chosen the arbitrary constant of integration $a_{\mathit{pt}}$ to parameterize the
time at which $K=1/2$, that is
\begin{equation}\label{eq:K soln}
K=\left[1+\left(\frac{a_{\mathit{pt}}}{a}\right)^6\right]^{-1},
\end{equation}
so that it indicates the time at which the gravitational and matter metrics are close to coinciding. The subscript $\mathit{pt}$ indicates that this is the ``phase transition'', when standard local Lorentz symmetry with a single light cone will be restored.

The Friedmann equation is given by
\begin{equation}\label{eq:reduced Fried eqn}
 H^2 +\frac{kc^2}{a^2} =\frac{1}{3}c^2\Lambda+H_B^2\left(\frac{a_{\mathit{pt}}}{a}\right)^6,
\end{equation}
At early enough times the behavior of the universe is dominated by the contribution from the biscalar field. We neglect the cosmological constant and spatial curvature throughout the remainder of this work. We have the approximate solution
\begin{equation}\label{eq:Rsoln}
a= a_{\mathit{pt}}(3H_Bt)^{1/3},\quad
H=\frac{1}{3t},
\end{equation}
where we have chosen $t=0$ when $a=0$, and at the phase transition $t_{\mathit{pt}}=1/(3H_B)$, which means that $H_{\mathit{pt}}=H_B$.

It appears that this is not an inflationary solution, so it does not solve the horizon and flatness problems. However, we are working in a comoving gravitational frame, and the speed of matter propagation is very much larger than that of gravitational waves. The coordinate distance that light would travel since the initial singularity is given by the null geodesics of~\eqref{eq:FRW ghat} rather than those of~\eqref{eq:gravmetric}. It follows that
\begin{equation}
\int \frac{c\,dt}{Ka}\propto \frac{1}{t^{4/3}},
\end{equation}
where we have used~\eqref{eq:Rsoln}. This clearly diverges as $t\rightarrow 0$, showing that there is no (matter) particle horizon in the spacetime and the horizon problem is resolved.
Similar arguments can be employed to solve the flatness problem~\cite{Albrecht+Maguiejo:1999}.

In the comoving gravitational frame with a minimally-coupled Einstein Klein-Gordon field, the scalar mode fluctuations: $\phi=\phi(t)+\delta\phi(t,\vec{x})$ about the background cosmological solution can be treated in a straightforward way. We find the field equation for the perturbation of the biscalar field:
\begin{equation}\label{eq:phi pert}
\frac{d^2\delta\phi_{\vec{k}}}{dt^2}+3H\frac{d \delta\phi_{\vec{k}}}{dt}+\frac{c^2\vec{k}^2}{a^2}\delta\phi_{\vec{k}}=0,
\end{equation}
where 
\begin{equation}
\delta\phi  = (2\pi)^{-2/3}\int d^3k\,\exp[-i\vec{k}\cdot\vec{x}]\delta\phi_{\vec{k}}.
\end{equation}
The solution to~\eqref{eq:phi pert} for our spacetime is given by Bessel functions:
\begin{equation}
\delta\phi_{\vec{k}} = A(\vec{k})J_0(\xi)+B(\vec{k})Y_0(\xi),\quad
\xi=\frac{ck}{2a_{\mathit{pt}}H_B}\left(\frac{a}{a_{\mathit{pt}}}\right)^2.
\end{equation}
Since $\xi\gtrsim 1$ would correspond to a mode passing inside the horizon, we expect that in the very early universe we would have $\xi\ll 1$ for modes of interest.
Using the fact that $a_0H_0/(a_{\mathit{pt}}H_{\mathit{pt}})\ll 1 $, for modes near the pivot point: $ck\sim 7 a_0H_0$ we see that this is the case.
Thus, we cannot assume a scenario where the modes are ``born'' in the quantum vacuum in the early universe.

Rewriting~\eqref{eq:phi pert} in terms of the matter comoving time $\tau$ results in a field equation that is very different than in what would appear in a deSitter spacetime:
\begin{equation}
\frac{d^2\delta\phi_{\vec{k}}}{d\tau^2}
+\frac{d\ln\left(a^3/\sqrt{K}\right)}{d\tau}\frac{d \delta\phi_{\vec{k}}}{d\tau}
+\frac{Kc^2\vec{k}^2}{a^2}\delta\phi_{\vec{k}}=0.
\end{equation}
The additional factor $K$ multiplying the spatial derivative terms is a reflection of the fact that the null cone of the biscalar field is smaller by exactly this factor over that of matter.
Since in the very early universe $K\sim a^6$, the biscalar field modes never appear to be ``inside the horizon''.

Hollands and Wald~\cite{Hollands+Wald:2002a} assume that modes are born or emerge in the ground state of a flat spacetime wave operator at some length scale $\ell_0$. The modes emerge in a semi-classical state as a normalized plane wave of the flat spacetime wave operator:
\begin{equation}
\label{eq:classicalwave}
\delta\phi_{\vec{k}}(t_k) = \sqrt{\frac{\kappa \hbar c^2}{(2\pi a_k)^32\omega_k}}
\cos(\omega_k t_k -\vec{k}\cdot\vec{x} +\delta),\quad\omega_k =\frac{ck}{a_k},
\end{equation}
where the scale at which the mode is born is given by the condition:
\begin{equation}\label{eq:HW condition}
a_k=k\ell_0.
\end{equation}

We match not only the initial perturbation but also its time derivative.  Keeping only the dominant contribution as $\xi\rightarrow 0$ gives
\begin{equation}
\delta\phi_{\vec{k}}\approx\sqrt{\frac{9\kappa \hbar c^2}{(2\pi a_k)^3 32\omega_k}}\cos(\omega_k t_k -\vec{k}\cdot\vec{x} +\delta)\ln(\xi_k) J_0(\xi),
\end{equation}
where $\xi_k$ represents $\xi$ evaluated at $a=a_k$. There is a non-scale invariant contribution, for we matched to the initial state and its time derivative, and the Bessel function $Y_0(z)$ is logarithmically divergent when $\xi\rightarrow 0$. This will lead to slight deviations from a scale invariant spectrum.

We obtain the spectrum of scalar field fluctuations to be
\begin{equation}
\mathcal{P}_{\delta\phi}
=\frac{9}{2(2\pi)^3}\left(\frac{\ell_P}{\ell_0}\right)^2
\ln^2(\xi_k).
\end{equation}
The curvature power spectrum is then given by (recall that
$\dot{\phi}^2=12 H^2$ in this spacetime):
\begin{equation}\label{eq:PR}
\mathcal{P}_{\mathcal{R}}
=\biggl(\frac{H^2}{\dot{\phi}^2}\biggr)\mathcal{P}_{\delta\phi}
=\frac{3}{8(2\pi)^3}\left(\frac{\ell_P}{\ell_0}\right)^2
\ln^2(\xi_k),
\end{equation}
where by using the condition~\eqref{eq:HW condition} we have
\begin{equation}\label{eq:xi k}
\xi_k=\frac{\sqrt{12B}\ell^2_0 k^3}{2a_{\mathit{pt}}^3}.
\end{equation}

From~\eqref{eq:PR} the scalar mode spectral index is calculated:
\begin{equation}
n_s=1+\frac{d\ln \mathcal{P}_{\mathcal{R}}}{d\ln k}=1+\frac{6}{\ln(\xi_k)},
\end{equation}
and the running of the spectral index is calculated from: $\alpha_s=dn_s/d\ln k$, from which we find the relation
\begin{equation}
\alpha_s=-\tfrac{1}{2}(1-n_s)^2.
\end{equation}
In the large scale limit, we also have
\begin{equation}\label{eq:delta H}
\delta_H= \frac{2}{5}\sqrt{\mathcal{P}_{\mathcal{R}}}
\approx
\frac{2}{5}\sqrt{\frac{3}{8(2\pi)^3}}\biggl(\frac{\ell_P}{\ell_0}\biggr)
|\ln(\xi_k)|.
\end{equation}
Assuming that $\ell_0\approx \sqrt{12B}$ and evaluating the logarithm at the pivot point, $c k \sim 7 R_0H_0$, this simplifies to
\begin{equation}
\delta_H\approx \frac{\ell_P}{\ell_0} .
\end{equation}
To see this, note that for this scale: $\xi_k\approx (7 H_0R_0/H_{\mathit{pt}}R_{\mathit{pt}})^3$, where we have used: $H_{\mathit{pt}}=H_B$ and~\eqref{eq:definitions}.
Using standard physics following $R_{\mathit{pt}}$, one finds that $H_{\mathit{pt}}R_{\mathit{pt}}/R_0 H_0\approx \sqrt{10^{60} z_{\mathrm{eq}}}(T_P/T_{\mathrm{pt}})$~\cite{Coles+Lucchin:1995}, which gives $\ln(\xi_k)\approx - (150 -200)$, which combines with the coefficient in~\eqref{eq:delta H} to give a factor of order unity. We can then match the amplitude of the observed CMB fluctuations: $\delta_H\sim 10^{-5}$ by fixing the bimetric length scale: 
\begin{equation}\label{eq:l0 constraint}
\ell_0\approx \sqrt{12B} \approx 10^5 \, \ell_P,
\end{equation}
which gives
\begin{equation}
n_s \approx 0.96,\quad
\alpha_s \approx -4.5 - -8 \times 10^{-4}.
\end{equation}
This compares well with the recent Planck satellite mission result: $n_s=0.96\pm 0.0073$~\cite{PlanckMission1,PlanckMission2,PlanckMission3}. While it is straightforward to use this mechanism to also generate a nearly scale invariant spectrum of tensor fluctuations, we expect that as with the scalar mode fluctuations the gravitational or tensor mode fluctations will be outside the horizon, so the ratio $r={\rm tensor}/{\rm scalar}$ will be zero. The assumption that we can take $\bar{g}_{\mu\nu}\approx g_{\mu\nu}$ can be justified by showing that $\kappa c^2 B \sqrt{K}\rho \ll 1$ since from~\eqref{eq:K soln}, $K\ll 1$ in the very early universe. 

A VSL motivated scheme based on a scalar-tensor bimetric gravity theory leads successfully to an approximately scale invariant CMB spectrum with $n_s\approx 0.96$. Moreover the model predicts that the ratio  $r={\rm tensor}/{\rm scalar}$ will be zero. This prediction distinguishes the model from inflationary models in which $r$ is expected to be non-zero. The model does not suffer from many of the fine-tuning problems of standard inflation models. The fluctuations that yield a spectral index in agreement with the Planck2013 mission result do not suffer from the production of large inhomogeneities in the later universe, which in the chaotic and eternal inflation models are rationalized in terms of a multiverse and anthropic principle scenario with a resulting lack of predictability. 

\section*{acknowledgments}

This research was generously supported by the John Templeton Foundation. Research at the Perimeter Institute for Theoretical Physics is supported by the Government of Canada through NSERC and by the Province of Ontario through the Ministry of Research and Innovation (MRI).

\end{document}